# Voids and Mn-rich inclusions in a (Ga,Mn)As ferromagnetic semiconductor investigated by transmission electron microscopy

A. Kovács[1*], J. Sadowski[2,3], T. Kasama[1], J. Domagała[3], R. Mathieu[4], T. Dietl[3,5] and R. E. Dunin-Borkowski[1]

[1] Center for Electron Nanoscopy, Technical University of Denmark, Kgs. Lyngby 2800, Denmark

[2] MAX-Lab, Lund University, P.O. Box 118, 221 00 Lund, Sweden

[3] Institute of Physics, Polish Academy of Sciences, al. Lotników 32/46, 02-668 Warszawa, Poland

[4] Department of Engineering Sciences, Uppsala University, P.O. Box 534, SE-751 21 Uppsala, Sweden

[5] Institute of Theoretical Physics, University of Warsaw, PL-00-681 Warszawa, Poland

Voids adjacent to both cubic (ZnS-type) and hexagonal (NiAs-type) Mn-rich nanocrystals are characterized using aberration-corrected transmission electron microscopy in annealed $Ga_{0.995}Mn_{0.005}As$ magnetic semiconductor specimen grown by molecular beam epitaxy. Nano-beam electron diffraction measurements suggest that the nanocrystals exhibit deviations in lattice parameter from that of bulk MnAs. *In situ* annealing inside the electron microscope is used to study the nucleation, coalescence, and grain growth of individual nanocrystals. After annealing at 903 K, the magnetic transition temperature of the specimen likely to be dominated by the presence of cubic ferromagnetic nanocrystals.

## INTRODUCTION

Magnetic semiconductors are materials that can exhibit both ferromagnetic (FM) and semiconducting properties, and are widely studied because of their intriguing physical properties and potential applications in spin-based electronic devices.[1] It has become clear in recent years[2] that the detailed identification of the atomic arrangement in such materials is indispensible for understanding their basic properties, particularly the origin of ferromagnetism. For (Ga,Mn)As that has been deposited at low temperatures, typically below 573 K, the Mn ions are randomly distributed over cation[3] and interstitial sites.[4] However, in the case of the post-growth annealing[5] or growth[6] at higher temperatures, the Mn ions tend to aggregate into Mn-rich (Ga,Mn)As ferromagnetic nanocrystals buried in the GaAs lattice. Such nanocrystals have been found to exhibit the cubic (ZnS-type) structure of the host for annealing temperatures of below 773 K (Refs. 7-9). In contrast, precipitation of hexagonal (NiAs-type) ferromagnetic nanocrystals has been observed for higher annealing temperatures.[5,7-9] The appearance of both chemical and crystallographic phase separation is a generic property of magnetically doped semiconductors, which results from a sizeable contribution of open magnetic shells to the cohesive energy.[2]

For the particular semiconductor/ferromagnet nanocomposite system GaAs:MnAs, remarkable functionalities have recently been demonstrated, including enhanced magnetooptical[10] and magnetotransport[6] properties, the direct conversion of magnetization energy into electric current,[11] and an extra long spin-relaxation times in the Coulomb blockade regime.[12] A wealth of other functionalities,[13,14] has been predicted for this and related systems, awaits for an experimental verification.

The characterization of small nanocrystals that contain aggregated transition metal ions using transmission electron microscopy (TEM) is highly challenging, because their size and their arrangement can differ little from that of the host. Although conventional bright-field phase contrast imaging has previously been used to study Mn-rich nanocrystals in GaAs layers,[5, 7-9] no detailed investigations have so-far been carried out. Recent improvements in spatial resolution and chemical sensitivity of TEM techniques have resulted from the development of aberration correctors[15,16] for both TEM and scanning TEM (STEM) methods.[17] Annular dark-field (ADF) imaging in the STEM, in particular, is sensitive to variations in atomic number $Z$ and strain, as well as permitting the acquisition of electron energy-loss spectra simultaneously with the ADF signal.[18]

Here, we use aberration-corrected TEM

and aberration-corrected STEM ADF imaging to examine heat-treated $Ga_{0.995}Mn_{0.005}As$ epilayers. We show that both cubic and hexagonal Mn-rich nanocrystals can be associated with closely adjacent voids. We determine the lattice parameters of the nanocrystals using nano-beam electron diffraction (NBD). *In situ* annealing of the layers inside of the microscope is used to suggest that the nanocrystals and voids may form independently. The local Mn distribution in individual nanocrystals is studied using electron energy-loss spectroscopy (EELS) and the magnetic properties of the annealed layers are discussed.

**EXPERIMENTAL**

(Ga,Mn)As layers were grown on GaAs (001) substrates by using molecular beam epitaxy (MBE) in a KRYOVAK system.[19] A 30-nm-thick high temperature GaAs buffer layer was deposited prior to (Ga,Mn)As growth. The substrate temperature was then decreased to 543 K and 700 nm of $Ga_{0.995}Mn_{0.005}As$ was grown. $As_2$ dimmers were supplied from a DCA valve cracker source, operating at a stage temperature of 1173 K. The $As_2$/Ga flux ratio during (Ga,Mn)As growth was close to 2. After MBE growth, the samples were taken out of the vacuum system, cleaved into several pieces and reintroduced into the MBE growth chamber for annealing in ultra-high vacuum conditions at moderate (673 K) and high (833 and 903 K) temperatures for 60 min. Here, we concentrate on the samples that were annealed at 903 K.

Structural characterization and chemical analysis were performed on cross-sectional TEM specimens prepared using conventional mechanical polishing and Ar ion milling. The specimens were finished using low ion energies (< 1 keV) in order to minimize ion beam induced sample preparation artifacts. Both image aberration-corrected and probe aberration-corrected TEM and STEM studies were carried out using FEI Titan microscopes operated at 300 kV. The inner semi-angle of the ADF detector was varied between 24 and 98 mrad when collecting both low-angle ADF (LAADF) and high-angle ADF (HAADF) signals. Information about the Mn distribution was obtained using a combination of STEM ADF images and EELS line-scans. A highly parallel electron beam with a 2.6 nm full-width at half maximum was used for NBD experiments. Simulated diffraction patterns were obtained using the Java version of EMS software. Thermal annealing studies were carried out by *in situ* in the microscope using a heating holder. Electron diffraction patterns, STEM ADF and bright-field (BF) TEM images were used to follow the structural and chemical changes in the specimen upon the annealing. Magnetic properties of the samples were studied using a superconductive quantum interference device (SQUID) magnetometer.

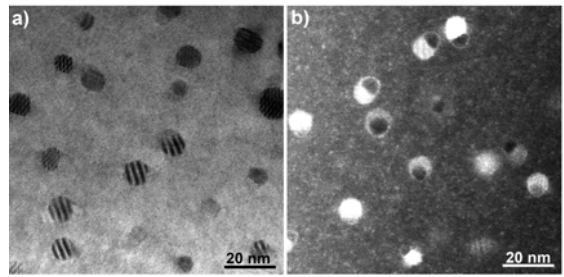

FIG. 1. Low magnification (a) bright-field and (b) LAADF STEM images of (Ga,Mn)As annealed at 903 K. The ADF inner detector semi-angle used was 24.5 mrad. The viewing direction is close to the crystallographic [1-10] axis of the GaAs host lattice.

**RESULTS AND DISCUSSION**

Figure 1 shows representative low magnification BF TEM and LAADF STEM images of the sample that had been annealed at 903 K. In the TEM image shown in Fig. 1(a), Moiré fringes are visible at the positions of many of the nanocrystals, which are approximately equidimensional and have an average size of 10.8 nm. The nanocrystals are identified to have either cubic (ZnS-type, space group 216) or hexagonal (NiAs-type, space group 194) structures from NBD studies (see below). Surprisingly, in the LAADF image (Fig.1 (b)) almost all of the nanocrystals have dark regions adjacent to them, which we interpret as voids. The voids, which have bright rims around them in the LAADF image, do not appear to have preferential locations or sizes with respect to the nanocrystal orientations, structures or sizes.

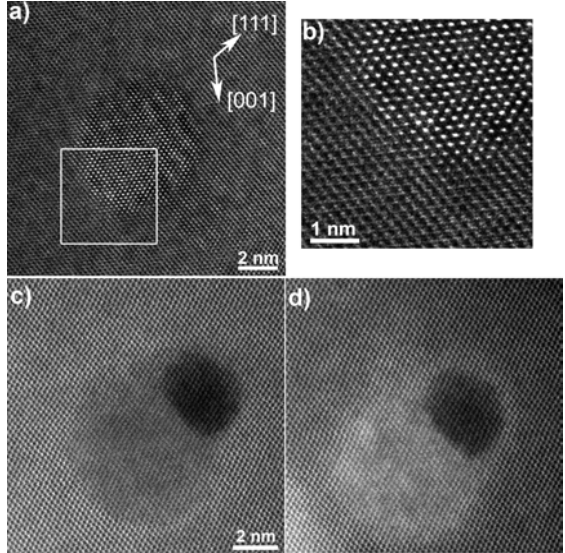

FIG. 2. High-resolution aberration-corrected (a, b) TEM, (c) HAADF STEM, and (d) LAADF STEM images of cubic (Ga,Mn)As nanocrystals. The ADF detector inner semi-angles used are (c) 78.4 and (d) 24.5 mrad.

Figure 2 shows aberration-corrected high-resolution images (both TEM and ADF STEM) of two different cubic (Ga,Mn)As nanocrystals. In the TEM images shown in Figs. 2(a) and 2(b), the objective lens aberrations were corrected up to fourth order and a small negative value of the spherical aberration coefficient $C_s$ (-3.5 μm) was used. At a defocus of approximately -32 nm, the GaAs host and the (Ga,Mn)As nanocrystals were observed to exhibit different contrast. The enlargement in Fig. 2(b) confirms that the nanocrystal is fully coherent with the GaAs matrix and that no dislocations are present. In such an image, an adjacent void would be barely visible. Figure 2(c) shows a probe-aberration corrected STEM HAADF image of a different nanocrystal. The presence of lighter Mn atoms ($Z_{Mn}$= 25, $Z_{Ga}$= 31, $Z_{As}$= 33) in the nanocrystal results in slightly darker contrast than that of the GaAs host, whereas the adjacent much darker region is a void. Interestingly, just as in Fig. 1(b) a bright band of contrast is visible around the void in an LAADF image acquired from the same nanocrystal, as shown in Fig. 2(d). Yu et al.[18] showed that for a thick (> 15 nm) Si sample strain contrast can result in bright contrast in LAADF images and dark contrast in HAADF images. In our studies, bands of bright contrast in LAADF images are present around both cubic and hexagonal nanocrystals and have widths of 1-1.5 nm.

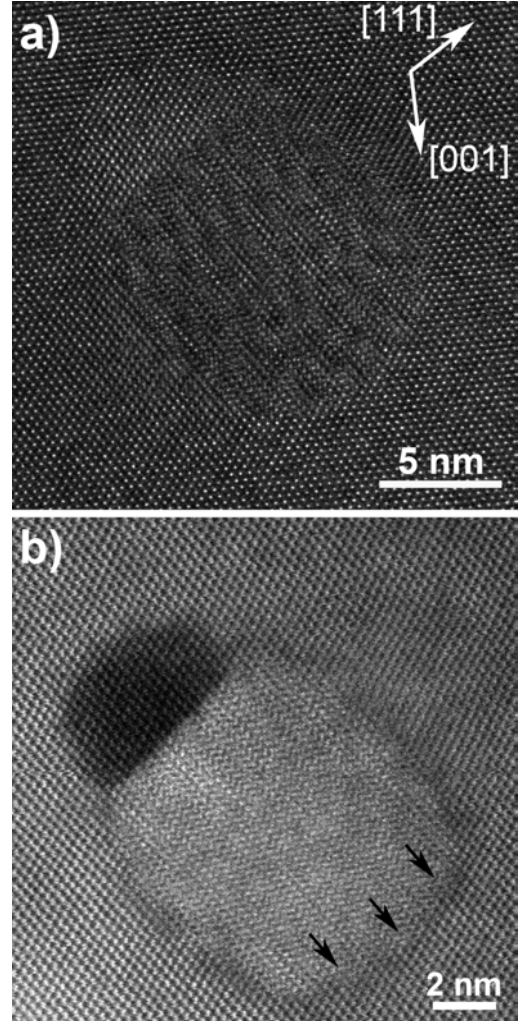

FIG. 3. Aberration-corrected high-resolution (a) TEM and (b) HAADF STEM images of two different hexagonal (Ga,Mn)As nanocrystals. The arrows indicate the position of misfit dislocations. In (b), the inner detector semi-angle was used was 78.4 mrad.

Figure 3 illustrates void formation adjacent to two different hexagonal nanocrystals. The overlapping lattices complicate the high-resolution TEM image shown in Fig. 3(a). In contrast, in the corresponding HAADF image shown in Fig. 3(b), the hexagonal lattice can be resolved, with misfit dislocations visible at the interface between the nanocrystal and the GaAs, while the void forms a sharp interface with the nanocrystal. In Fig. 3(b) a dark band of contrast is visible around the nanocrystal. Although similar contrast was not as prominent around the cubic crystal shown in Fig. 2(c), the difference may result from the brighter contrast of the hexagonal nanocrystal shown in Fig. 3(b), relative to that of the host GaAs lattice. The bright contrast of the hexagonal crystal may be associated with either diffraction contrast or a difference in Mn concentration when compared with the cubic crystal shown in Fig. 2(c). The fact that the band around the crystal is dark in both cases suggests that it is associated with strain around the nanocrystals.

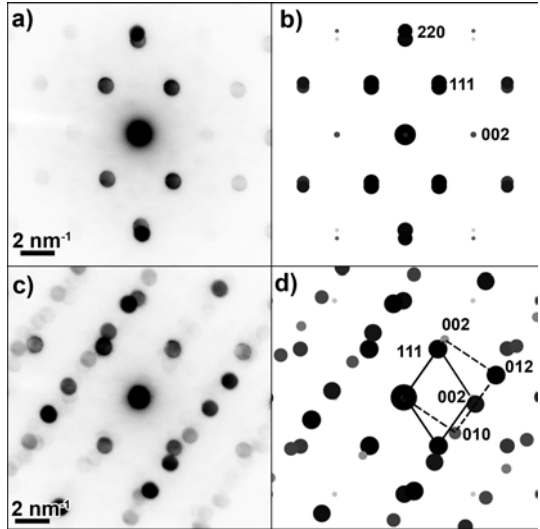

FIG. 4. (a,c) Experimental and (b, d) simulated electron diffraction patterns of (a,b) cubic and (c, d) hexagonal (Ga,Mn)As nanocrystals in GaAs. The viewing direction is [1-10] for GaAs.

NBD technique was used to determine the structures and lattice parameters of individual cubic and hexagonal nanocrystals. Representative experimental and simulated diffraction patterns of cubic and hexagonal nanocrystals with diameters of 8 and 10 nm are shown in Fig. 4. The measured lattice parameter of the cubic crystal matches that of the host GaAs lattice in the (001) growth direction (Figs. 4(a) and 4(b)), while a decrease in lattice parameters is observed in the orthogonal direction. The inferred lattice parameters for the nanocrystal of $c = 0.565$ nm and $a = 0.489 \pm 0.005$ nm are consistent with the Moiré fringes observed in Fig. 1(a) and with the separation of the 220 reflections in the diffraction pattern shown in Fig. 4(a). Despite the lattice mismatch of -13 % with respect to GaAs, the crystal lattice of the nanocrystal matches that of the GaAs host, as revealed in the TEM images shown in Fig. 2. In contrast, crystallographic relationship between the hexagonal nanocrystals and the GaAs lattice is determined to be $[1\text{-}10]_{GaAs}//[2\text{-}1\text{-}10]_{hex.}$, $(111)_{GaAs}//(0002)_{hex.}$. There are therefore four possible orientations of the hexagonal $c$-axis (which is the magnetic hard axis of MnAs) with respect to {111} planes of GaAs. By using both diffraction patterns and Fourier transforms patterns of TEM lattice images (not shown), the lattice parameters of the hexagonal structure were determined to be $a(b) = 0.359$ nm and $c = 0.589 \pm 0.005$ nm. For comparison, the bulk values that are $a = 0.372$ nm and $c = 0.5713$ nm,[20] corresponding to -3.5 and +3.1 % misfits in the $a$ and $c$ directions, respectively.

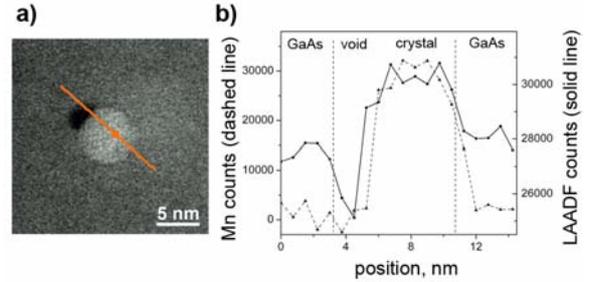

FIG. 5. (a) LAADF (inner detector semi-angle 47.4 mrad) STEM image of a hexagonal (Ga,Mn)As nanocrystal. (b) LAADF intensity and Mn $L$ edge EELS signal collected after background subtraction along the line marked in (a).

LAADF and EELS line-scans were recorded simultaneously in order to measure the Mn compositional profile across a hexagonal (Ga,Mn)As nanocrystal. Figure 5 shows a plot of the intensity integrated under the Mn $L$ edge (640 eV), collected point-by-point across a void-nanocrystal combination in ~0.75 nm steps, plotted alongside the corresponding LAADF intensity profile. The dip in the LAADF profile corresponds to the position of the void. The Mn signal is below to detection limit of the present measurements in both the GaAs host and the void, and increases when the electron beam reaches the nanocrystal. Interestingly, the LAADF signal increases before the Mn signal on entering the nanocrystal from the direction of the void. The origin of this difference is not understood at present. However, it suggests that the crystal-void interface may have a different composition to the interior of the crystal.

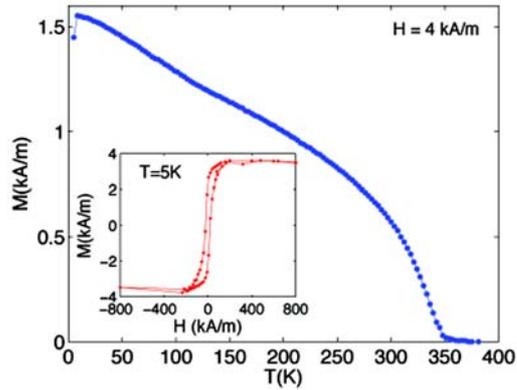

FIG. 6. Temperature dependence of the field-cooled magnetization for $Ga_{0.995}Mn_{0.005}As$ annealed at 903 K recorded in a magnetic field of 4 kA/m (50 Oe). The inset shows an hysteresis loop at T=5 K (corrected by subtracting a diamagnetic contribution).

Complementary SQUID magnetometry results acquired from the same sample, which are shown in Fig. 6, are consistent with the presence of cubic Mn-rich nanocrystals, which have a higher Curie temperature ($T_C$=350 K) than that with a NiAs-type hexagonal structure ($T_C$=313 K).[5-8,19] Although we observed both structures in the samples that had been annealed at 833 K and 903 K, the temperature for magnetization onset in field-cooled measurements is determined by the magnetic nanocrystals that have the higher value of $T_C$.

The leading mechanism accounting for the void formation adjacent to the cubic and hexagonal nanocrystals is not fully understood. It is known that the relatively low growth temperature that is needed to incorporate a sizeable concentration of Mn in GaAs (applied in this study, too) also results in a high density of point defects, of which the most important are As anti-sites and Mn interstitials.[4] The annihilation of these defects during high temperature annealing may result in the formation of Mn-rich nanocrystals and vacancies. Furthermore, the process of Mn aggregation proceeds, presumably, *via* the generation of Mn vacancy – Mn interstitial pairs. Hence, heat treatment creates vacancies, whose accumulation may than lead to the formation of voids. It is interesting to note that the presence of voids in semiconductor light emitting diodes (LEDs), by affecting optical properties, may serve to increase the efficiency of LEDs.[21]

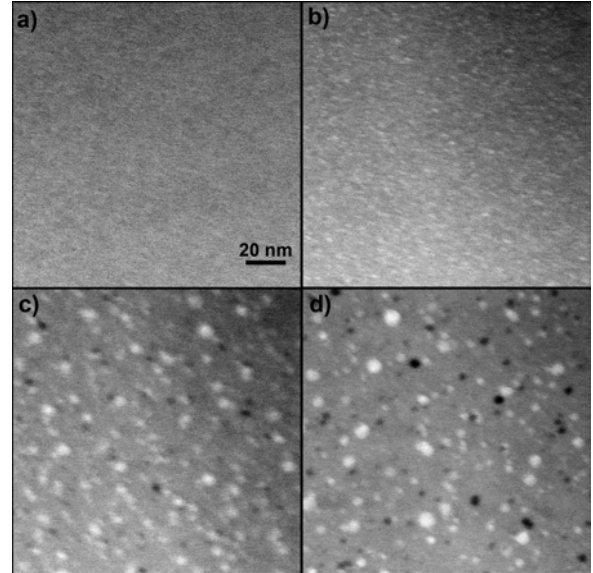

FIG. 7. LAADF (inner detector semi-angle 47.4 mrad) STEM images recorded at indicated specimen temperatures of (a) 773 K, (b) 823 K, (c) 848 K and (d) 848 K 6 minutes later.

In order to further understand the nature of void formation and Mn aggregation in GaAs, we annealed an as-grown $Ga_{0.995}Mn_{0.005}As$ sample inside the microscope and recorded LAADF images of the resulting microstructural and chemical changes, as shown in Fig. 7. The contrast of the layer in a cross-sectional specimen was uniform for annealing temperatures of up to 773 K [Fig. 7(a)]. At a temperature of approximately 823 K, the formation of nanocrystals with sizes of 1 to 4 nm was observed, as shown in Fig. 7(b). The nanocrystals, which exhibited bright contrast in the LAADF images, were observed to coalesce and grow to larger sizes when the temperature was increased to 848 K (Fig. 7(c)). Void formation was then also observed at positions that did not appear to be related to the positions of the nanocrystals. When the specimen was held at this temperature, the voids became larger without changing their locations, while the nanocrystals moved, coalesced and grew to larger sizes (Fig. 7(d)). The final morphology of the sample that had been annealed *in situ* in the microscope was clearly different from the observed in the sample that had been annealed in ultra-high vacuum conditions in the MBE chamber. This difference may result from the fact that the *in situ* annealing was carried out in different kinetic conditions. However, it illustrates the complexity of the processes that are involved and highlights the need for further dedicated studies of such systems under realistic conditions of elevated temperature and pressure in the TEM.

## CONCLUSIONS

Our results indicate that, despite many reports devoted to the GaAs:MnAs nanocomposite system and to its high temperature post-growth annealing, the formation of Mn-rich nanocrystals with cubic and hexagonal structures is not yet fully understood. ADF STEM images reveal void formation adjacent to both cubic and hexagonal MnAs nanocrystals in GaAs host during *ex situ* and *in situ* annealing at temperatures of up to 903 K. Bands of contrast that may be associated with strain are observed around both the nanocrystals and the voids using ADF imaging. *In situ* heating experiments in the microscope suggest that the nanocrystals and the voids may form independently, with the nanocrystal sizes, shapes and positions evolving over time during annealing and voids remaining more static. Our results also suggest that the onset of the ferromagnetic properties of the annealed (Ga,Mn)As layers is determined by the presence of the cubic rather than the hexagonal nanocrystals.

## ACKNOWLEDGEMENT
This work was supported by the "FunDMS" Advanced Grant of the European Research Council within the "Ideas" 7th Framework Programme of the European Commission.